\documentclass[%
 reprint,
 amsmath,amssymb,
 aps,
citeautoscript]{revtex4-2}

\setcitestyle{super}

\usepackage[normalem]{ulem}
\usepackage{xcolor}

\usepackage{amsmath}
\usepackage{tikz}
\usetikzlibrary{matrix}

\usepackage{amssymb}
\usepackage{amsthm}
\usepackage{graphicx}
\usepackage{dcolumn}
\usepackage{bm}
\usepackage{upgreek} 
\usepackage[colorlinks=true,bookmarks=false,citecolor=black,urlcolor=blue,linkcolor=black]{hyperref} 

\usepackage{color}
\usepackage{soul}
\soulregister\cite{7} 
\soulregister\ref{7} 
\soulregister\eqref{7} 

\usepackage{xcolor}

\begin{document}

\title{Mode Switching Through Exceptional Points Induced by Lasing–Inversion Coupling}

\author{Xingwei Gao$^1$} \email{xingweig@utexas.edu}
\author{Cheng Guo$^1$}
\author{David Burghoff$^1$}

\affiliation{
$^1$Chandra Department of Electrical and Computer Engineering, University of Texas at Austin, TX 78712, USA
}

\begin{abstract}
\section*{Abstract}
The gain–loss coupling in optical cavities induces exceptional points (EPs), where two optical modes coalesce. The large modal overlap near an EP intensifies gain competition, favoring single-mode lasing. Recent studies further revealed self-modulation closer to the EP that transforms the lasing mode into a frequency comb. Such EP-enabled comb formation suggests a previously unaccounted-for mechanism that overcomes the strong gain competition and drives a second mode to threshold.
Here, using a Bloch coupled-mode theory derived from first principles, we show that the second threshold arises from dynamical couplings among the population inversion, the lasing field, and a dark cavity mode. The lasing–inversion coupling produces extra EPs, whose spectral structure governs switching among single-mode lasing and frequency combs with different repetition rates. This above-threshold mode-switching mechanism enables new opportunities for tunable photonic systems, including adaptive optical communication links and dual-comb spectroscopy.
\end{abstract}

\maketitle

\section{Introduction}\label{Introduction}

\vspace{-4pt}   
An exceptional point (EP) is a non-Hermitian spectral degeneracy at which both the eigenfrequencies and the associated modal fields coalesce~\cite{2011_Moiseyev_book,Heiss_2012,feng2017non,el2018non,doi:10.1126/science.aar7709}. 
EPs arise from the interplay of gain and loss, making lasers a natural platform for their realization in optical systems. In EP laser cavities, the collapse of modal orthogonality reshapes the eigenvalue landscape and modifies the spatial structure of the resonant fields, giving rise to unconventional laser dynamics, including chiral emission~\cite{doi:10.1073/pnas.1603318113,doi:10.1126/science.aaf8533,doi:10.1126/science.aba8996}, spectral phase transitions~\cite{2021_Roy_ncomms}, and topological state transfer~\cite{schumer2022topological}. 
When one cavity mode near an EP reaches threshold, the other mode is strongly suppressed due to the gain competition intensified by their large modal overlap. 
This nonlinear gain saturation, combined with operation in the parity-time-symmetry-broken regime, makes EP lasers promising candidates for robust single-mode lasing~\cite{doi:10.1126/science.1258479,doi:10.1126/science.1258480}.

More intriguingly, when the spectral separation between the two modes becomes comparable to the inversion decay rate, the single-mode lasing state can lose stability~\cite{PhysRevA.96.053837}, causing the EP laser to spontaneously evolve into a frequency comb---a set of equally spaced spectral lines~\cite{PhysRevA.107.033509,ji2023,Gao2024PALT}. 
Optical frequency combs are conventionally generated either by external modulation~\cite{Parriaux:20,2007_DelHaye_Nature,2018_Kippenberg_Science_review} or through intrinsic self-modulation involving multiple active cavity modes~\cite{hugi2012mid,silvestri2023frequency,2024_Opacak_Nature}. 
The emergence of self-modulated EP combs therefore points to a distinct mechanism that circumvents the strong gain competition near EPs and enables a second mode to reach threshold following the onset of lasing. 
However, despite extensive numerical demonstrations~\cite{PhysRevA.96.053837,PhysRevA.107.033509,ji2023,Gao2024PALT}, a clear physical explanation for the near-EP mode boosting remains lacking.

To understand this unconventional mode behavior, we develop a Bloch coupled-mode theory (B-CMT) for gain–loss coupled cavities, which provides a quantitatively accurate reduced description derived from first-principles laser theory.
Within the B-CMT framework, we show that the lasing field dynamically couples population-inversion fluctuations to the optical modes through a Bogoliubov--de Gennes Hamiltonian $\mathbf{H}_{\mathrm{BdG}}$, originally introduced in superconductivity theory~\cite{gor1958energy,1999Gennes}.
The eigenmodes of $\mathbf{H}_{\mathrm{BdG}}$
can be activated by the intracavity lasing field and subsequently drive self-modulation that generates frequency combs.
$\mathbf{H}_{\mathrm{BdG}}$ includes additional EPs arising from the coupling between the lasing field and the inversion.
 {Their} spectral topology determines which mode reaches threshold and the required pumping strength, thereby enabling controlled switching between single-mode lasing and frequency combs with different repetition rates.
This EP-enabled mode-switching mechanism opens new opportunities for EP lasers in applications such as dual-comb spectroscopy and adaptive optical communication.
\begin{figure*}[t]
\centering
\includegraphics{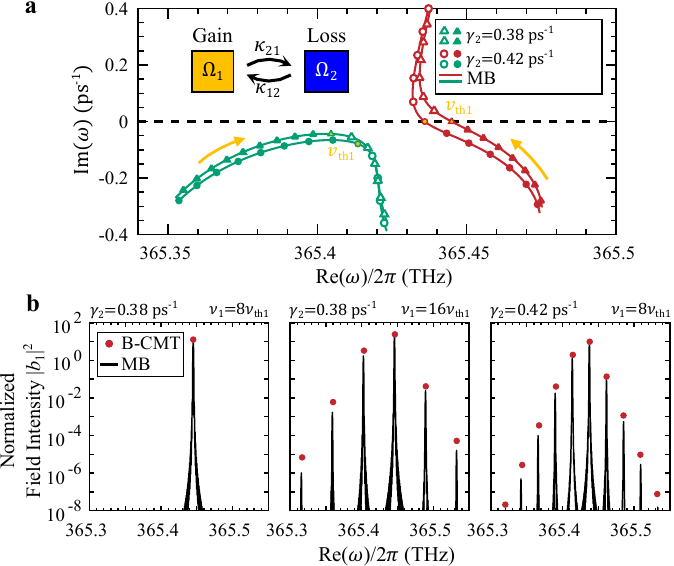}
\vspace{-4pt}
\caption{\textbf{Laser behavior of gain--loss coupled cavities near an exceptional point.} 
\textbf{a} Eigenfrequency trajectories of the two optical modes $\mathrm{O_I}$ (red) and $\mathrm{O_{II}}$ (green), obtained from $\mathbf{H}_0(\nu_1)$, for two values of the passive cavity loss. Symbols denote B-CMT results, while solid lines show Maxwell--Bloch simulations. Yellow arrows indicate increasing pump strength $\nu_1$, and $\nu_{\mathrm{th}1}$ marks the first lasing threshold. Inset: schematic of the gain--loss coupled-cavity system. 
\textbf{b} Lasing spectra above the first threshold for different passive losses and pumping strengths, showing the transition from a fixed point (single-mode lasing, left panel) to limit cycles (frequency combs, middle and right panels). The field intensity is normalized by the factor $f=\hbar\sqrt{\gamma_\perp \gamma_\parallel}/(2R)$. The values of all system parameters are included in Supplementary section~II.
}
\vspace{-4pt}
\label{fig_CMT_MB}
\end{figure*}
   
\section{Results}
Second-order EPs in lasers are most commonly realized through gain–loss compensation between two linearly coupled cavities~\cite{feng2017non, el2018non, PhysRevLett.122.093901}, as illustrated schematically in the inset of Fig.~\ref{fig_CMT_MB}\textbf{a}. The two uncoupled cavities are characterized by complex resonant frequencies $\Omega_1$ and $\Omega_2$, with passive losses $\gamma_{1,2} = -\mathrm{Im}(\Omega_{1,2})$, while optical gain is introduced into cavity~1 through external pumping with strength $\nu_1$. By appropriately tuning the gain, loss, and coupling coefficients $\kappa_{12,21}$, the system can be brought close to an EP at the first lasing threshold $\nu_1=\nu_{\mathrm{th}1}$, where two eigenfrequencies nearly coalesce and one crosses the real axis, as shown in Fig.~\ref{fig_CMT_MB}\textbf{a}.

In this regime, the conventional linear-gain picture predicts a repelling phase of the EP above threshold: once lasing begins, the two eigenfrequencies separate predominantly along the imaginary axis, strongly suppressing the nonlasing partner mode. This behavior implies robust single-mode lasing, corresponding to a stable fixed point in the language of nonlinear dynamics (black curve in Fig.~\ref{fig_CMT_MB}\textbf{b}, left panel). Such enhanced mode competition is generally regarded as a defining feature of EP lasers and underlies their appeal as intrinsically single-mode sources.

Surprisingly, full-wave time-domain simulations reported in our previous work~\cite{Gao2024PALT, Gao_nph} revealed that this expectation can break down above threshold. As the pump strength is increased, the fixed point loses stability and the system spontaneously evolves into a limit cycle, producing a frequency-comb spectrum (middle panel of Fig.~\ref{fig_CMT_MB}\textbf{b}). This transition signals the activation of a second frequency component that dynamically couples to the lasing mode, enabling differential-frequency generation despite the strong EP-enhanced mode competition. Even more counter-intuitively, increasing the passive loss $\gamma_2$ of the non-active cavity can further promote this instability, leading to combs with smaller repetition rates and richer spectral structure (right panel of Fig.~\ref{fig_CMT_MB}\textbf{b}).

These simulation results were initially obtained by numerically solving the Maxwell–Bloch equations~\cite{haken1985laser, PhysRevA.54.3347}, a microscopic density-matrix framework that self-consistently couples electromagnetic fields to the gain medium. While Maxwell–Bloch simulations faithfully capture the full laser dynamics, their explicit spatial dependence and strong nonlinearity render the resulting solutions highly non-analytic and difficult to interpret physically. Moreover, the absence of an effective modal decomposition of the nonlinear dynamics obscures the origin of the second frequency component and the mechanism by which it overcomes EP-induced mode competition. A reduced yet quantitatively accurate description is therefore essential for isolating the relevant dynamical degrees of freedom and uncovering the physical origin of this unexpected above-threshold behavior.

\subsection{Bloch coupled-mode theory}\label{sec:BCMT} 
For systems composed of coupled single-mode cavities, the spatial structure of limit-cycle solutions can be systematically averaged using quasi-normal modes and a Pad\'{e}-approximant reduction~\cite{He2025}. 
Here we extend this approach to Maxwell--Bloch equations in the time domain and obtain a Bloch coupled-mode theory describing the interaction among single-mode cavities of arbitrary geometry (see Supplementary Section~I for the derivation),
\small
\begin{align}
\dot d_1(t) &= - \gamma_{\parallel}(d_1-\nu_{1}) - \tfrac{2i}{\hbar}\beta_1[b_1^*p_1-\mathrm{c.c.}], \label{BCMT:d}\\
\dot{p}_1(t) &= -(i\omega_a + \gamma_{\perp})p_1 - \frac{iR^2}{\hbar}d_1b_1, \label{BCMT:p}\\
i\dot{b}_1(t) &= \Omega_1 b_1 +\kappa_{12} b_2 + \varepsilon_0^{-1} C_{11}\ddot{p}_1, \label{BCMT:b1} \\
i\dot{b}_2(t) &= \Omega_2 b_2 +\kappa_{21} b_1, \label{BCMT:b}
\end{align}
\normalsize
Here $d_1 \in \mathbb{R}$ denotes the population inversion in the active cavity, $p_1$ the inversion-induced polarization, and $b_1$ and $b_2$ the optical-field amplitudes in the active and passive cavities, respectively. The parameters $\omega_a$, $\gamma_{\perp}$, and $\gamma_{\parallel}$ are the gain center frequency, polarization decay rate, and population relaxation rate. The coefficient $\beta_1 \in \mathbb{R}$, obtained from Pad\'e fitting, is determined by the spatial mode profile of the active cavity, while $\nu_1$ represents the effective pumping strength. The coupling between the polarization and the cavity field is described by $C_{11}$. Here $R$ is the atomic dipole moment amplitude, $\varepsilon_0$ the vacuum permittivity, and $\hbar$ the reduced Planck constant. In the narrow-band lasing regime, Eqs.~\eqref{BCMT:d}--\eqref{BCMT:b} reduce to conventional rate-equation models in which the nonlinear gain depends on the intracavity intensity $|b_1|^2$~\cite{2017_Teimourpour_srep, 2018_Kominis_APL, ji2023, PhysRevLett.130.266901}. The general Bloch temporal coupled-mode theory for an arbitrary number of active and passive cavities is presented in Supplementary Section~I.

At or below the lasing threshold, the system resides at the trivial steady state
$
[d_1,p_1,b_1,b_2]^\mathrm{T}=[\nu_1,0,0,0]^\mathrm{T}
$. 
Linearizing Eqs.~(\ref{BCMT:d})--(\ref{BCMT:b}) about this state yields a $4\times4$ Jacobian matrix $-i\mathbf{H}_0$, where
\small
\begin{equation}
\mathbf{H}_0(\nu_1)=
\begin{bmatrix}
 -i\gamma_\parallel & 0 & 0 & 0 \\
 0 & \omega_a-\omega_0-i\gamma_\perp & \dfrac{R^2}{\hbar}\nu_1 & 0 \\
 0 & -\dfrac{C_{11}}{\varepsilon_0}\omega_0^2 & \Omega_1-\omega_0 & \kappa_{12} \\
 0 & 0 & \kappa_{21} & \Omega_2-\omega_0
\end{bmatrix}.
\label{H0}
\end{equation}
\normalsize
The first lasing threshold $\nu_{\mathrm{th}1}$ and the corresponding lasing frequency $\omega_0\in\mathbb{R}$ are simultaneously determined by the condition $\det[\mathbf{H}_0(\nu_{\mathrm{th}1})]=0$.

Under this condition, one of the four eigenvalues of $\mathbf{H}_0(\omega_0,\nu_1)$, denoted as $\omega_{\mathrm{I}}$, reaches zero at $\nu_1=\nu_{\mathrm{th}1}$ and corresponds to the lasing mode labeled $\mathrm{O}_{\mathrm{I}}$. A second eigenvalue and its corresponding eigenmode are labeled $\omega_{\mathrm{II}}$ and $\mathrm{O}_{\mathrm{II}}$, respectively. These two optical modes continuously connect to the eigenmodes of the passive system $\mathbf{H}_{\mathrm{L}}$ at $\nu_1=0$. Their eigenfrequency trajectories, $\omega_{\mathrm{I}}(\nu_1)+\omega_0$ and $\omega_{\mathrm{II}}(\nu_1)+\omega_0$, characterize the laser behavior under linear gain and show excellent agreement with Maxwell--Bloch simulations (Fig.~\ref{fig_CMT_MB}\textbf{a}).

A third eigenmode of $\mathbf{H}_0$ corresponds to population-inversion fluctuations with a fixed decay rate $\gamma_\parallel$. Because $\mathbf{H}_0$ is block diagonal, this mode is completely decoupled from $\mathrm{O}_{\mathrm{I}}$ and $\mathrm{O}_{\mathrm{II}}$. Above the lasing threshold, however, nonlinear gain saturation couples inversion fluctuations to the optical modes, a mechanism that will be analyzed in Sec.~\ref{sec:LIH}. The remaining eigenmode stays near $\omega_a - i\gamma_\perp$ and represents fast polarization relaxation, which plays no significant role in the lasing dynamics of interest.

Above threshold, numerical integration of Eqs.~\eqref{BCMT:d}--\eqref{BCMT:b} also reproduces the transition from fixed points to limit cycles observed in Maxwell--Bloch simulations (Fig.~\ref{fig_CMT_MB}\textbf{b}). In the B-CMT model, the spatial profiles of $\mathrm{O}_{\mathrm{I}}$ and $\mathrm{O}_{\mathrm{II}}$ within the gain cavity are assumed to be identical, since only one mode from each individual cavity is retained in the reduced description. The quantitative agreement between B-CMT and Maxwell--Bloch simulations therefore rules out spatial hole burning and additional cavity modes separated by the free spectral range as the physical origin of the limit cycles observed in Fig.~\ref{fig_CMT_MB}\textbf{b}.

\begin{figure*}[t]
\centering
\includegraphics{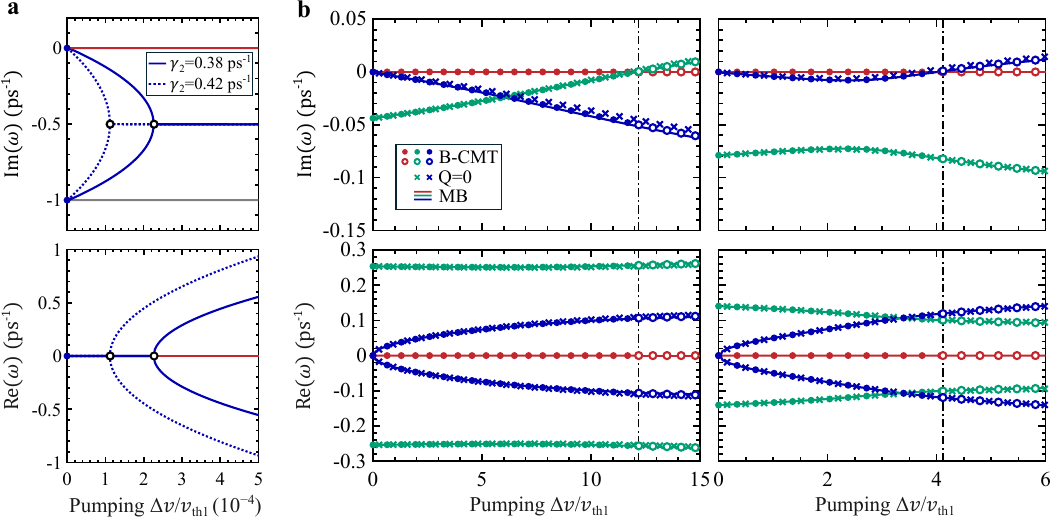}
\vspace{-4pt}
\caption{\textbf{Lasing--inversion hybrid (LIH) modes above the first threshold.} 
\textbf{a} The eigenvalue spectrum of $\mathbf{H}_{\mathrm{BdG}}$ for small relative pumping $\Delta\nu$. The lasing mode pinned at $\omega=0$ (red), the non-self-conjugate inversion fluctuation pinned at $-i\gamma_\parallel$ (gray), the $\mathrm{LIH}_{\mathrm{I}}$ modes (blue), and the Hamiltonian Krein-Hopf Bifurcation (black circle) are shown. 
\textbf{b} Eigenvalue trajectories of the $\mathrm{LIH}_{\mathrm{I}}$ (blue) and $\mathrm{LIH}_{\mathrm{II}}$ (green) modes as functions of pumping $\Delta\nu$ beyond the bifurcation. The left column corresponds to $\gamma_2=0.38~\mathrm{ps}^{-1}$, while the right column corresponds to $\gamma_2=0.42~\mathrm{ps}^{-1}$. Circles denote eigenvalues of $\mathbf{H}_{\mathrm{BdG}}$ given by Eq.~\eqref{HBdG}, while crosses indicate the roots of the quartic equation $Q(\omega,\Delta\nu)=0$ defined in Eq.~\eqref{Q}.
}
\vspace{-4pt}
\label{fig_LIH}
\end{figure*} 

\subsection{Lasing-inversion hybrid modes}\label{sec:LIH}
Above the first lasing threshold, the system resides at a single-mode fixed point stabilized by gain saturation. At this fixed point, the population inversion is clamped at $d_1=\nu_{\mathrm{th}1}$, while the polarization and optical fields oscillate harmonically at the lasing frequency $\omega_0$, namely $p_1=P_1 e^{-i\omega_0 t}$ and $b_{1,2}=B_{1,2} e^{-i\omega_0 t}$. Substituting these steady-state solutions into Eqs.~\eqref{BCMT:d}--\eqref{BCMT:b1}, one finds that the intracavity intensity increases linearly with pump above threshold,
\begin{equation}
    |B_1|^2
    =\frac{\hbar^2\gamma_\perp\gamma_\parallel}{4R^2\beta_1|\Gamma_\perp|^2}
    \frac{\Delta\nu}{\nu_{\mathrm{th}1}},
\end{equation}
where $\Delta\nu\equiv\nu_1-\nu_{\mathrm{th}1}$ and $\Gamma_\perp=\gamma_\perp/(\omega_0-\omega_a+i\gamma_\perp)$.

Although the inversion is stationary at the fixed point, it is not dynamically inert. A small fluctuation of the population inversion, $\delta D_1(t)$, arising for example from spontaneous emission or environmental perturbations, enters the polarization equation~\eqref{BCMT:p} as a source term proportional to $B_1\delta D_1(t)$. This perturbation modulates the polarization amplitude, $p_1\rightarrow [P_1+\delta P_1(t)]e^{-i\omega_0 t}$, which in turn radiates additional optical fields, $b_{1,2}(t)\rightarrow [B_{1,2}+\delta B_{1,2}(t)]e^{-i\omega_0 t}$. Conversely, fluctuations of the polarization and optical fields feed back onto the inversion via the light--matter interaction (the second term on the right-hand side of Eq.~\eqref{BCMT:d}), introducing contributions proportional to $B_1^*\delta P_1(t)$ and $P_1\delta B_1^*(t)$ in the inversion dynamics. As a result, inversion, polarization, and optical-field fluctuations are mutually generated and dynamically coupled around the fixed point.

To describe this coupling quantitatively, we define the perturbation vector
\[
\bm{\delta}(t)\equiv[\delta D_1,\,\delta P_1,\,\delta B_1,\,\delta B_2]^T,
\]
and linearize Eqs.~\eqref{BCMT:d}--\eqref{BCMT:b} about the fixed point. This yields
\begin{equation}
    i\frac{d}{dt}
    \begin{bmatrix}
        \bm{\delta} \\
        \bm{\delta}^*
    \end{bmatrix}
    =\mathbf{H}_{\mathrm{BdG}}
    \begin{bmatrix}
        \bm{\delta} \\
        \bm{\delta}^*
    \end{bmatrix},
\end{equation}
where $\mathbf{H}_{\mathrm{BdG}}$ is an $8\times8$ Bogoliubov--de Gennes (BdG) Hamiltonian~\cite{gor1958energy,1999Gennes},
\small
\begin{equation}
    \mathbf{H}_{\mathrm{BdG}}=
    \begin{bmatrix}
        \mathbf{H}_0(\nu_{\mathrm{th}1}) & \mathbf{0} \\
        \mathbf{0} & -\mathbf{H}_0^*(\nu_{\mathrm{th}1})
    \end{bmatrix}
    +\frac{|B_1|}{\hbar}
    \begin{bmatrix}
        \mathbf{\Delta} & \mathbf{K} \\
        -\mathbf{K}^* & -\mathbf{\Delta}^*
    \end{bmatrix}.
    \label{HBdG}
\end{equation}
\normalsize
Here $\mathbf{\Delta}$ and $\mathbf{K}$ are $4\times4$ sparse matrices encoding the lasing-field–induced coupling between inversion and optical fluctuations, with nonzero elements
\small
\begin{align}
    &\mathbf{\Delta}_{12}=2\beta_1, \qquad
    \mathbf{\Delta}_{13}=-\frac{2\beta_1\Gamma_\perp^*R^2\nu_{\mathrm{th}1}}{\hbar\gamma_\perp}, \qquad
    \mathbf{\Delta}_{21}=R^2, \\
    &\mathbf{K}_{12}=-\mathbf{\Delta}_{12}^*, \qquad
    \mathbf{K}_{13}=-\mathbf{\Delta}_{13}^* .
\end{align}
\normalsize

The derivation of Eq.~\eqref{HBdG} is provided in Supplementary Section~III. The eigenvalue spectrum of $\mathbf{H}_{\mathrm{BdG}}$ is symmetric with respect to the imaginary axis. Following the convention in superconductivity physics, we refer to this symmetry as particle--hole symmetry (PHS), although no charge carriers are involved in the present optical system.

The global $U(1)$ symmetry of Eqs.~\eqref{BCMT:d}--\eqref{BCMT:b} ensures that the lasing mode $\mathrm{O}_{\mathrm{I}}$ is an eigenvector of $\mathbf{H}_{\mathrm{BdG}}$ with an eigenvalue fixed at $0$ (red line in Fig.~\ref{fig_LIH}\textbf{a}). A second eigenvalue is pinned at $-i\gamma_\parallel$ (gray line in Fig.~\ref{fig_LIH}\textbf{a}), corresponding to a non-self-conjugate (and therefore non-physical) inversion fluctuation 
(see Supplementary Section~III). 


At the first lasing threshold, where the lasing-field amplitude $|B_1|$ vanishes, $\mathbf{H}_{\mathrm{BdG}}$ reduces to two decoupled blocks, $\mathbf{H}_0(\nu_{\mathrm{th}1})$ and $-\mathbf{H}_0^*(\nu_{\mathrm{th}1})$. In this limit, the optical mode $\mathrm{O}_{\mathrm{I}}$ and its PHS partner are degenerate at $\omega=0$ in the complex spectrum, while the physical and non-self-conjugate inversion-fluctuation modes are degenerate at $-i\gamma_\parallel$. Above threshold, the emergence of a finite lasing field couples inversion fluctuations to $\mathrm{O}_{\mathrm{I}}$ through the matrices $\mathbf{\Delta}$ and $\mathbf{K}$, lifting these degeneracies. As the pump strength increases, PHS symmetry constrains the resulting LIH modes to evolve along the imaginary axis, collide at an exceptional point, and subsequently split along the real axis. 
This symmetry-enforced spectral transition constitutes a Hamiltonian Krein--Hopf bifurcation~\cite{marsden2013introduction}, as illustrated in Fig.~\ref{fig_LIH}\textbf{a}. We refer to the PHS-related mode pair emerging from this bifurcation as $\mathrm{LIH}_{\mathrm{I}}$, with eigenvalues $(\tilde{\omega}_{\mathrm{I}}, -\tilde{\omega}_{\mathrm{I}}^{*})$.

The remaining pair of BdG modes, denoted as $\mathrm{LIH}_{\mathrm{II}}$, arises from the coupling between the population inversion and the optical mode $\mathrm{O}_{\mathrm{II}}$. Their eigenvalues are denoted as $(\tilde{\omega}_{\mathrm{II}},-\tilde{\omega}_{\mathrm{II}}^{*})$. At the first threshold, $\tilde{\omega}_{\mathrm{II}}=\omega_{\mathrm{II}}$.

Fig.~\ref{fig_LIH}\textbf{b} shows $\tilde{\omega}_{\mathrm{I}}$ (blue) and $\tilde{\omega}_{\mathrm{II}}$ (green) as functions of the pump strength beyond the Hamiltonian Krein-Hopf Bifurcation. For a passive cavity loss of $\gamma_2=0.38~\mathrm{ps}^{-1}$, $\mathrm{LIH}_{\mathrm{I}}$ remains inhibited, while $\tilde{\omega}_{\mathrm{II}}$ crosses the real axis. Beyond this point, the inversion fluctuation associated with $\mathrm{LIH}_{\mathrm{II}}$ modulates the system, yielding the limit cycle shown in the middle panel of Fig.~\ref{fig_CMT_MB}\textbf{b}. When the passive loss is increased to $\gamma_2=0.42~\mathrm{ps}^{-1}$, the roles of the two LIH modes are reversed: $\mathrm{LIH}_{\mathrm{I}}$ turns on and generates a distinct limit cycle, as shown in the right panel of Fig.~\ref{fig_CMT_MB}\textbf{b}. This mode-switching behavior is governed by another exceptional point, which will be discussed in Sec.~\ref{sec:LIH-EP}.

\subsection{LIH--EP}\label{sec:LIH-EP}
We show that the LIH eigenvalues of $\mathbf{H}_{\mathrm{BdG}}$ can be accurately approximated by the roots of a quartic polynomial $Q(\omega,\Delta\nu)$ that preserves PHS (see Supplementary Section~IV),
\begin{equation}
\begin{aligned}
        Q(\omega,\Delta\nu) &= \omega(\omega+i\gamma_\parallel)
        (\omega-\omega_{\mathrm{II}})(\omega+\omega_{\mathrm{II}}^{*}) \\
        &\quad - \frac{\gamma_\perp \gamma_\parallel}{2|\Gamma_\perp|^2}
        \left(s_2\omega^2 - i s_1\omega - s_0\right),
        \label{Q}
\end{aligned}
\end{equation}
where $s_0$, $s_1$, and $s_2\in\mathbb{R}$ are pump-independent coefficients determined by the profiles of the optical modes $\mathrm{O}_{\mathrm{I}}$ and $\mathrm{O}_{\mathrm{II}}$ at the first threshold. The solutions of $Q(\omega,\Delta\nu)=0$ show excellent agreement with both the B-CMT model and the full MB equations, as shown in Fig.~\ref{fig_LIH}\textbf{b}.

\begin{figure*}[t]
\centering
\includegraphics{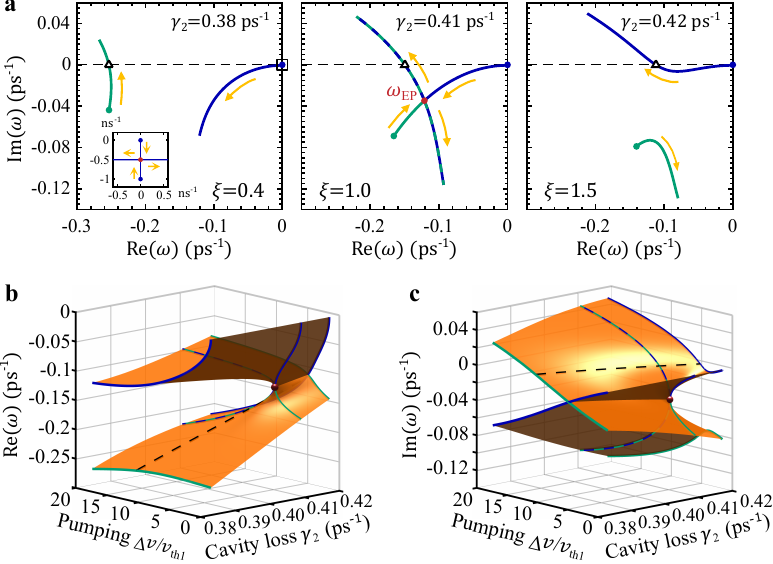}
\caption{\textbf{Mode switching through LIH-induced exceptional points.} 
\textbf{a} Trajectories of the LIH eigenvalues $\tilde{\omega}_{\mathrm{I}}$ (blue) and $\tilde{\omega}_{\mathrm{II}}$ (green) obtained from the quartic equation $Q(\omega,\Delta\nu)=0$ for different values of the passive cavity loss $\gamma_2$. Triangles indicate crossings of the eigenvalues with the real axis, marking the second lasing threshold. The red dot in the middle panel denotes the LIH-EP frequency $\omega_{\mathrm{EP}}$ satisfying Eq.~\eqref{LIH-EP}. The inset in the left panel shows a zoomed-in view of $\tilde{\omega}_{\mathrm{I}}$ near threshold, highlighting the PHS-EP bifurcation shown in Fig.~\ref{fig_LIH}\textbf{a}. Yellow arrows indicate increasing pump strength $\Delta\nu$. 
\textbf{b} Real and \textbf{c} imaginary parts of the LIH eigenvalue surfaces as functions of $\gamma_2$ and $\Delta\nu$. Blue and green curves correspond to the $\gamma_2$ slices shown in \textbf{a}. The red sphere marks the LIH-EP, and the black dashed line denotes the second lasing threshold defined by $\mathrm{Im}(\omega)=0$.
}
\label{fig_LIH_O2_EP}
\end{figure*}

A second exceptional point, referred to as the LIH-EP, occurs at $\gamma_2=0.41~\mathrm{ps}^{-1}$, where the eigenvalue trajectories $\tilde{\omega}_{\mathrm{I}}$ and $\tilde{\omega}_{\mathrm{II}}$ intersect (middle panel of Fig.~\ref{fig_LIH_O2_EP}\textbf{a}). Within the quartic effective model, the LIH-EP corresponds to a degeneracy of the form
\begin{equation}
    Q(\omega)=(\omega-\omega_{\mathrm{EP}})^2(\omega+\omega_{\mathrm{EP}}^{*})^2, \label{LIH-EP}
\end{equation}
which yields the condition $\xi=1$, where
\begin{equation}
    \xi =
    \frac{4s_0\,\mathrm{Im}(\omega_{\mathrm{II}})}
    {s_1\,\mathrm{Re}(\omega_{\mathrm{II}})^2}
    \left(1-\frac{s_2}{s_1}\mathrm{Im}(\omega_{\mathrm{II}})\right),
\end{equation}
and $\omega_{\mathrm{II}}=\omega_{\mathrm{II}}(\nu_1=\nu_{\mathrm{th1}})$ is the eigenvalue of  {the} $\text{O}_\text{II}$ mode at the first threshold. 

The dimensionless parameter $\xi$ distinguishes two distinct phases away from the LIH-EP. When $\xi<1$, the LIH-EP splits along the imaginary axis (left panel of Fig.~\ref{fig_LIH_O2_EP}\textbf{a}), forcing $\mathrm{LIH}_{\mathrm{I}}$ to move downward while $\mathrm{LIH}_{\mathrm{II}}$ moves upward and crosses the lasing threshold. This explains the activation of the $\mathrm{LIH}_{\mathrm{II}}$ mode at $\gamma_2=0.38~\mathrm{ps}^{-1}$ and the resulting limit cycle shown in Fig.~\ref{fig_CMT_MB}\textbf{b}. 
When $\xi>1$, the LIH-EP splits along the real axis (right panel of Fig.~\ref{fig_LIH_O2_EP}\textbf{a}), enhancing $\mathrm{LIH}_{\mathrm{I}}$ while suppressing $\mathrm{LIH}_{\mathrm{II}}$. This accounts for the activation of the $\mathrm{LIH}_{\mathrm{I}}$ mode at $\gamma_2=0.42~\mathrm{ps}^{-1}$ and the corresponding limit cycle in Fig.~\ref{fig_CMT_MB}\textbf{b}. 
The observed mode-switching behavior is enabled by the Riemann-sheet structure of the eigenvalue surfaces around the LIH-EP in the $(\gamma_2,\Delta\nu)$ parameter space, as illustrated in Fig.~\ref{fig_LIH_O2_EP}\textbf{b} and \ref{fig_LIH_O2_EP}\textbf{c}.

\section{Conclusions}
In this work, we developed a quantitatively accurate Bloch temporal coupled-mode theory derived from the Maxwell--Bloch equations, which faithfully reproduces full time-domain simulations while providing a reduced and physically transparent description. Within this framework, we showed that population-inversion fluctuations couple dynamically to the lasing field above threshold, giving rise to LIH modes described by a Bogoliubov--de Gennes Hamiltonian. We demonstrated that the formation of LIH modes is governed by PHS-enforced Hamiltonian Krein-Hopf Bifurcation, and that their evolution controls the second threshold and the emergence of limit cycles. By further introducing a quartic effective model, we analytically identified an LIH-induced EP and revealed a mode-switching mechanism enabled by the associated Riemann-sheet structure.

The second threshold can be substantially reduced by increasing the passive-mode loss, a counterintuitive strategy that simultaneously enhances frequency-comb formation, yielding a larger number of comb lines and a broader spectral bandwidth \cite{Gao2024PALT, Gao_nph}. The ability to controllably switch between distinct LIH modes and their associated dynamical states opens new possibilities for reconfigurable frequency-comb sources, nonlinear signal generation, and mode-selective dynamical control in non-Hermitian photonic systems.

\vspace{8pt}
\noindent {\bf Acknowledgments}

\vspace{3pt}
\noindent {\bf Author contributions}
X.G. developed the B-CMT  {framework}, performed the eigenvalue analysis, and carried out the Maxwell--Bloch simulations. 
X.G. and C.G. developed the BdG analysis. D.B. supervised the project. X.G. wrote the manuscript with input 
from the other coauthors. All authors discussed the results.

\vspace{3pt}
\noindent {\bf Competing interests}
The authors declare no competing interests.

\vspace{3pt}
\noindent {\bf Data availability}
All data that support the findings of this study are presented in the paper and in the supplementary materials. 


\textbf{Supplementary Material}: This article contains supplementary material.

\bibliography{sn-bibliography}
\bibliographystyle{naturemag}


\end{document}


\title{Supplementary Information}
\maketitle
\onecolumngrid
\tableofcontents

\thispagestyle{plain} 

\section{Derivation of Bloch Coupled-Mode Theory}
Here we derive the B-CMT equations (1)–(3) in the main text from the Maxwell–Bloch (MB) equations~\cite{haken1985laser, PhysRevA.54.3347}, which govern the dynamics of the electric field $\mathbf{E}(\mathbf{r},t)$, the pump-induced polarization $\mathbf{P}(\mathbf{r},t)$, and the population inversion $D(\mathbf{r},t)$ inside the gain medium. To simplify the notation, in this derivation we normalize the variables as
\begin{CEquation}
D \to \frac{\varepsilon_0 \hbar \gamma_\perp}{R^2} D, 
~~ \mathbf{P} \to \frac{\varepsilon_0\hbar\sqrt{\gamma_\perp \gamma_\parallel}}{2R} \mathbf{P}, 
~~ \mathbf{E} \to \frac{\hbar\sqrt{\gamma_\perp \gamma_\parallel}}{2R}\mathbf{E}, \label{norm}    
\end{CEquation}
where $\gamma_\parallel$ denotes the population relaxation rate and $\gamma_\perp$ the polarization dephasing rate. $R$ is the amplitude of the atomic dipole moment. Under this normalization, the MB equations take the form
\begin{CAlign}
&\frac{\partial}{\partial t}D = - \gamma_{\parallel}(D-D_\text{p})-\frac{i\gamma_{\parallel}}{2}(\mathbf{E}^*\cdot\mathbf{P}-\mathbf{E}\cdot\mathbf{P}^*), \label{MB:D}\\
&\frac{\partial}{\partial t}\mathbf{P} = -(i\omega_{a} + \gamma_{\perp})\mathbf{P}-i\gamma_{\perp}D(\mathbf{E}\cdot \bm{\theta})\bm{\theta}^*, \label{MB:P} \\
-&\nabla\times\nabla\times \mathbf{E} - \frac{\varepsilon_c}{c^2} \frac{\partial^2}{\partial t^2}\mathbf{E} = \frac{1}{c^2}\left(\frac{\sigma}{\varepsilon_0} \frac{\partial}{\partial t}\mathbf{E}+ \frac{\partial^2}{\partial t^2}\mathbf{P} \right).  \label{MB:E}
\end{CAlign}
The unit vector $\bm{\theta}$ represents the atomic dipole moment and satisfies $\bm{\theta}\cdot\bm{\theta}^*=1$. The linear optical properties of the laser system are described by the relative permittivity $\varepsilon_c(\mathbf{r})$ and the conductivity $\sigma(\mathbf{r})$ in Eq.~\eqref{MB:E}. In this work, $\varepsilon_c(\mathbf{r})$ is assumed to be frequency independent. Material dispersion can be incorporated by adding the corresponding time-derivative terms of $\mathbf{E}$ to the right-hand side of Eq.~\eqref{MB:E}.

We consider an optical system consisting of $N$ coupled cavities. Each cavity $n$ occupies a volume $V_n \subset \mathbb{R}^3$. Let $\{\mathbf{E}_n |n=1,2,\dots\}$ denote the eigenmodes of the system, which satisfy
\begin{CEquation}
-\nabla\times\nabla\times \mathbf{E}_n(\mathbf{r}) + \frac{\omega_n^2}{c^2} \varepsilon_c(\mathbf{r})\mathbf{E}_n(\mathbf{r}) = 0, \label{eigeq}   
\end{CEquation}
with outgoing boundary conditions. The total electric field can be expanded in terms of these modes as
\begin{CEquation}
    \mathbf{E}(\mathbf{r},t) \approx \sum_{n=1}^N a_n(t)\mathbf{E}_n(\mathbf{r}), \label{ModeExpansion_E}
\end{CEquation}
where $a_{n}(t)$ is the complex mode amplitude. Substituting Eq.~\eqref{eigeq} and Eq.~\eqref{ModeExpansion_E} into Eq.~\eqref{MB:E} and applying the slowly varying envelope approximation, $\ddot{a}_n \approx \omega_n^2a_n-2i\omega_n \dot{a}_n$, yields
\begin{CEquation}
    2\varepsilon_c \sum_{n=1}^N \omega_n \mathbf{E}_n(\mathbf{r}) \big[i\dot{a}_n(t)-\omega_n a_n(t)\big] 
    = \frac{\sigma}{\varepsilon_0}\, \partial_t\mathbf{E}(\mathbf{r},t) + \partial_t^2 \mathbf{P}(\mathbf{r},t). \label{wave_a}
\end{CEquation}

To extract $\dot{a}_n(t)$ from Eq.~\eqref{wave_a}, an orthogonality relation for the spatially diverging modes $\{\mathbf{E}_n\}$ is required. A commonly used approach is to employ quasi-normal modes~\cite{2022_Sauvan_OE_review}. Specifically, one selects a finite volume $V \supset (\cup_n V_n)$ enclosed by a scattering boundary and replaces the boundary condition with virtual perfectly matched layers in the far-field region $\mathbb{R}^3 \backslash V$,
\begin{CEquation}
    \varepsilon_c(\mathbf{r}) = \varepsilon_\text{PML}, \quad  
    \mu(\mathbf{r}) = \mu_\text{PML}, \quad  
    \mathbf{r} \in \mathbb{R}^3 \backslash V,
\end{CEquation}
where $\varepsilon_\text{PML}$ and $\mu_\text{PML}$ are diagonal $3 \times 3$ tensors. With an appropriate choice of attenuation factors in the perfectly matched layers, all radiative modes $\mathbf{E}_n$ can be mapped onto spatially confined quasi-normal modes. For convenience, we use the same notation $\mathbf{E}_n$ for these quasi-normal modes, which now satisfy 
\begin{CEquation}
    (\omega_n^2-\omega_m^2)\int_{\mathbb{R}^3} \mathbf{E}_m \cdot \varepsilon_c \mathbf{E}_n \, d^3r = 0. \label{orth}
\end{CEquation}
Equation~\eqref{orth} establishes the orthonormality relation
\begin{CEquation}
    (\mathbf{E}_m,\mathbf{E}_n) \equiv \int_{\mathbb{R}^3} \mathbf{E}_m \cdot \varepsilon_c \mathbf{E}_n d^3r = \delta(m-n), \label{orthnorm}
\end{CEquation}
Applying Eq.~\eqref{orthnorm} to Eq.~\eqref{wave_a}, we obtain the equation of motion for $a_n(t)$,
\begin{CEquation}
    2\omega_n\big[i\dot{a}_n(t)-\omega_n a_n(t)\big]
    = \frac{1}{\varepsilon_0}\, \partial_t \int_V \sigma \mathbf{E}_n \cdot \mathbf{E}\, d^3r
    + \partial_t^2 \int_V \mathbf{E}_n \cdot \mathbf{P}\, d^3r. \label{CMT_B1}
\end{CEquation}
The integration volume on the right-hand side is reduced from $\mathbb{R}^3$ to $V$, since both $\sigma(\mathbf{r})$ and $\mathbf{P}(\mathbf{r})$ vanish outside the laser system. 

Through the Bloch equations~\eqref{MB:D} and~\eqref{MB:P}, the polarization $\mathbf{P}$ depends nonlinearly on the total electric field $\mathbf{E}(\mathbf{r},t)$, which is spanned by all $\mathbf{E}_n$ in Eq.~\eqref{ModeExpansion_E}. Consequently, the second term on the right-hand side of Eq.~\eqref{CMT_B1} couples multiple $a_n$ components nonlinearly, making the equation of motion difficult to evaluate directly. Since the gain medium is confined within the cavities, it is more convenient to transform ${\mathbf{E}_n}$ into a set of individual-cavity modes ${\bm{\psi}_n(\mathbf{r})}$, with each $\bm{\psi}_n$ localized inside its respective cavity:
\begin{CEquation}
    \frac{\int_{V_{m\ne n}} |\bm{\psi}_n|^2 d^3r}{\int_{V_n} |\bm{\psi}_n|^2 d^3r} \approx 0. \label{LocalBasis}
\end{CEquation}
Such a basis transformation is possible under the condition of weak coupling between single-mode cavities, where the profiles of all ${\mathbf{E}_n}$ differ only by a coefficient inside each cavity. In this case, an $N\times N$ transformation matrix $\mathbf{M}$ can be obtained from Eq.~\eqref{LocalBasis}, such that
\begin{CEquation}
    [\bm{\psi}_1, \bm{\psi}_2, \dots, \bm{\psi}_N] \equiv [\mathbf{E}_1, \mathbf{E}_2, \dots, \mathbf{E}_N] \mathbf{M}.
\end{CEquation}
The total electric field can then be expanded in the new basis as
\begin{CEquation}
    \mathbf{E}(\mathbf{r},t) = \sum_{n=1}^N b_n(t)\bm{\psi}_n(\mathbf{r}). \label{ModeExpansion_psi}
\end{CEquation}
The relation between Eq.~\eqref{ModeExpansion_E} and Eq.~\eqref{ModeExpansion_psi} yields
\begin{CEquation}
    \bm{a} = \mathbf{M}\bm{b}, 
\end{CEquation}
with $\bm{a} = [a_1, \dots, a_N]^T$ and $\bm{b} = [b_1, \dots, b_N]^T$. Substituting this relation into Eq.~\eqref{wave_a} transforms it into
\begin{CEquation}
    i\big( 2\mathbf{M}^T \bm{\omega} \mathbf{M} + i\bm{\sigma} \big)\dot{\bm{b}}
    = 2\mathbf{M}^T \bm{\omega}^2 \mathbf{M}\bm{b} + \ddot{\bm{p}}, \label{tcmt_1}
\end{CEquation}
where $\bm{\omega} = \mathrm{diag}[\omega_1, \dots, \omega_N]$, and
\begin{CAlign}
    [\bm{p}]_m &= \int_{V} \bm{\psi}_m \cdot \mathbf{P}\, d^3r, \label{pt}\\
    [\bm{\sigma}]_{mn} &= \frac{1}{\varepsilon_0} \int_{V} \sigma \, (\bm{\psi}_m \cdot \bm{\psi}_n)\, d^3r.
\end{CAlign}



The remaining task is to determine $\bm{p}(t)$ using the Bloch equations, Eq.~\eqref{MB:D} and Eq.~\eqref{MB:P}. Hereafter, we denote the Fourier transform of any time-dependent variable $f(t)$ as $f_\omega$. In this notation, the Bloch equations in the frequency domain are expressed as
\begin{CAlign}
&{D_\omega} = 2\pi \delta (\omega)D_\text{p} + \tfrac{1}{2}\Gamma_\parallel(\mathbf{E}_\omega^\dagger * {\mathbf{P}_\omega} - {\mathbf{E}_\omega} * \mathbf{P}_\omega^\dagger), \label{FMB:D}\\
&{\mathbf{P}_\omega} = \Gamma_\perp (\omega)({\mathbf{E}_\omega}\cdot \bm{\theta}) * {D_\omega}\bm{\theta}^*, \label{FMB:P} 
\end{CAlign}
where “$*$” denotes convolution, $\Gamma_\parallel(\omega)=\gamma_\parallel/(\omega+i\gamma_\parallel)$, $\Gamma_\perp(\omega)=\gamma_\perp/(\omega-\omega_{a}+i\gamma_\perp)$, and ${(...)}^\dagger_\omega={(...)}^*_{-\omega}$.  

In performing the Fourier transform, we exclude the homogeneous solutions of $\mathbf{P}(\mathbf{r},t)$ and $D(\mathbf{r},t)$ that decay as $e^{-\gamma_\parallel t}$ and $e^{-\gamma_\perp t}$, respectively, since these only affect the system’s transient response near the initial time. All long-term dynamics are preserved in the frequency domain.  

Equation~\eqref{FMB:P} also implies
\begin{CEquation}
 \mathbf{P}_\omega^\dagger = \Gamma_\perp^\dagger({\mathbf{E}_\omega}\cdot \bm{\theta})^\dagger * {D_\omega}\bm{\theta}, \label{FMB:Pdagger}  
\end{CEquation}
where we have used $D_\omega^\dagger={D_\omega}$. Substituting Eq.~\eqref{FMB:P} and Eq.~\eqref{FMB:Pdagger} into Eq.~\eqref{FMB:D} gives a symbolic expression for ${D_\omega}$:
\begin{CEquation}
    {D_\omega}=2\pi D_p\{\delta(\omega)-0.5\Gamma_\parallel[({\mathbf{E}_\omega}\cdot\bm{\theta})^\dagger * \Gamma_\perp({\mathbf{E}_\omega}\cdot \bm{\theta})-({\mathbf{E}_\omega}\cdot\bm{\theta}) * \Gamma_\perp^\dagger({\mathbf{E}_\omega}\cdot \bm{\theta})^\dagger]\}*^{-1}\delta(\omega). \label{FMB:DE}
\end{CEquation}
Here, “$*^{-1}$” denotes the inverse of the convolution operator. For discrete-frequency solutions, such as limit cycles with period $2\pi/\omega_\text{d}$, convolution reduces to matrix multiplication. In this case, Eq.~\eqref{FMB:P} and Eq.~\eqref{FMB:DE} recover the theory of PALT \cite{Gao2024PALT}. Conversely, we may regard Eq.~\eqref{FMB:P} and Eq.~\eqref{FMB:DE} as PALT in the limit $\omega_\text{d} \to 0$, which allows us to compute $[{\bm{p}_\omega}]_n=\int \bm{\psi}_n\cdot{\mathbf{P}_\omega}\, d^3r$ using the approach in our previous work\cite{He2025}.  

Specifically, when Eq.~\eqref{LocalBasis} holds, each $\bm\psi_n$ has negligible spatial overlap with the others inside its cavity, so that
\begin{CEquation}
    {\mathbf{P}_\omega}({\mathbf{E}_\omega}) \approx {\mathbf{P}_\omega}([\bm{b}_\omega]_m \bm{\psi}_m),~ \text{for}~\mathbf{r}\in V_m.
\end{CEquation}
Hence, 
\begin{CEquation}
    [{\bm{p}_\omega}]_n=\int_V \bm{\psi}_n\cdot{\mathbf{P}_\omega} d^3r= \Gamma_\perp[\bm{b}_\omega]_m * [\bm{d}_\omega]_m, \label{FMB:p}
\end{CEquation}
where
\begin{CEquation}
    [\bm{d}_\omega]_m= \int_{V_m} (\bm{\psi}_m\cdot\bm{\theta})(\bm{\psi}_m\cdot\bm{\theta}^*)D_\omega d^3r. \label{FMB:d} 
\end{CEquation}
We substitute Eq.~\eqref{FMB:DE} into Eq.~\eqref{FMB:p}, then apply Pad\'{e} approximant\cite{baker1961pade}, which yields
\begin{CEquation}
    [\bm{d}_\omega]_m \approx 2\pi \nu_{m}\bigg\{\delta(\omega)-0.5\Gamma_\parallel\beta_m\bigg[{[\bm{b}_\omega}]_m^\dagger*(\Gamma_\perp [\bm{b}_\omega]_m)-[\bm{b}_\omega]_m*(\Gamma_\perp^\dagger {[\bm{b}_\omega}]_m^\dagger)\bigg]\bigg\}*^{-1}\delta(\omega), \label{PD:d}
\end{CEquation}
where the real coefficients $\nu_m$ and $\beta_m$ are determined numerically to minimize the error between Eq.~\eqref{PD:d} and Eq.~\eqref{FMB:d}. In this work, we demand the approximation to be exact when $\bm{b}_\omega=\bm{0}$. In this condition,
\begin{CAlign}
    \nu_m &= \int_{V_m} D_p (\bm{\psi}_m\cdot\bm{\theta})(\bm{\psi}_m\cdot\bm{\theta}^*) d^3r,
\end{CAlign}
and the global phase of $\bm{\psi}_m$ is chosen to satisfy $\nu_m\in\mathbb{R}$, so $\nu_m$ can be interpreted as the average pumping inside the cavity $m$. 


Taking the inverse Fourier transform of Eq.~\eqref{FMB:p} and Eq.~\eqref{PD:d}, we obtain the Bloch equations free of spatial coordinates:
\begin{CAlign}
&\dot d_m = - \gamma_{\parallel}[d_m(t)-\nu_{m}] - \tfrac{i\gamma_{\parallel}}{2}\beta_m[b_m^*(t)p_m(t)-b_m(t)p_m^*(t)], \label{MBTCMT:d}\\
&\dot{p}_m = -(i\omega_{a} + \gamma_{\perp})p_m - i\gamma_{\perp}d_m(t)b_m(t), \label{MBTCMT:p}
\end{CAlign}
with $m=1,2,\dots,N$. Finally, we revert the normalized variables $d_m$, $p_m$ and $\bm{b}$ back into physical units using the normalization factors for $D$, $\mathbf{P}$ and $\mathbf{E}$, respectively in Eq.~\eqref{norm}. Eq.~\eqref{tcmt_1}, \eqref{MBTCMT:d} and \eqref{MBTCMT:p} can then be rewritten as
\begin{CAlign}
&\dot d_m(t) = - \gamma_{\parallel}(d_m-\nu_{m}) - \tfrac{2i}{\hbar}\beta_m[b_m^*p_m-\mathrm{c.c.}], \label{BCMT:d}\\
&\dot{p}_m(t) = -(i\omega_a + \gamma_{\perp})p_m - \frac{iR^2}{\hbar}d_mb_m, \label{BCMT:p}\\
&i\dot{\bm{b}}=\mathbf{H}_\text{L}\bm{b}+\tfrac{1}{\varepsilon_0}\mathbf{C}\ddot{\bm{p}}, \label{BCMT:b}
\end{CAlign}
where
\begin{CAlign}
\mathbf{H}_\text{L} &= 2\mathbf{C}\mathbf{M}^T\bm{\omega}^2\mathbf{M}, \label{HL} \\
\mathbf{C} &= \big( 2\mathbf{M}^T \bm{\omega} \mathbf{M}+i\bm{\sigma} \big)^{-1}. \label{C}
\end{CAlign}
Eqs.~\eqref{BCMT:d}--\eqref{BCMT:b} are the Bloch coupled-mode theory. For two-cavity systems, we write 
$\mathbf{H}_\text{L}$ as
\begin{CEquation}
   \mathbf{H}_\text{L} = 
   \begin{bmatrix}
       \Omega_1 & \kappa_{12} \\
       \kappa_{21} & \Omega_2
   \end{bmatrix},
\end{CEquation}
and define cavity losses as $\gamma_{1,2}=-\text{Im}(\Omega_{1,2})$, then Eqs.~\eqref{BCMT:d}--\eqref{BCMT:b} become Eqs.~(1)--(3) in the main text. 

\section{Demonstration of B-CMT with a One-Dimensional Example} \label{1D-laser}
The B-CMT framework applies to couplings among single-mode cavities of arbitrary geometry and dimensionality, 
whereas direct finite-difference simulations of the MB equations near EPs are currently feasible only in 
one-dimensional systems. Therefore, for verification we benchmark B-CMT against the MB equations using a 
one-dimensional EP laser system from our previous work~\cite{Gao2024PALT}, which consists of two Fabry--P\'{e}rot 
cavities coupled through a distributed-feedback Bragg reflector.

For one-dimensional systems, the orthogonality relation in Eq.~\eqref{orth} can be enforced using analytical 
scattering boundaries rather than PMLs. Specifically, we consider fields of the form 
$\mathbf{E}(\mathbf{r},t)=\hat{y}E(x,t)$ with $\partial_y=\partial_z=0$, $\bm{\theta}=\theta\hat{y}$, and outgoing 
boundaries at $x=x_-$ and $x=x_+$. The corresponding scattering boundary condition is
\begin{CEquation}
    \partial_xE_n\big|_{x_\pm} = \pm i\frac{\Omega_n}{c}E_n(x_{\pm}). \label{1D_Boundary}
\end{CEquation}
Under this condition, the inner product in Eq.~\eqref{orthnorm} reduces to
\begin{CEquation}
    (\mathbf{E}_m,\mathbf{E}_n) = \int_{x_-}^{x_+} \varepsilon_c E_m E_n \, dx 
    + \frac{ic}{\Omega_m+\Omega_n}\big[ E_m(x_+)E_n(x_+) + E_m(x_-)E_n(x_-) \big]. \label{orthnm_1D}
\end{CEquation} 

\begin{figure}[h!]
\centering
\includegraphics[width=1.0\textwidth]{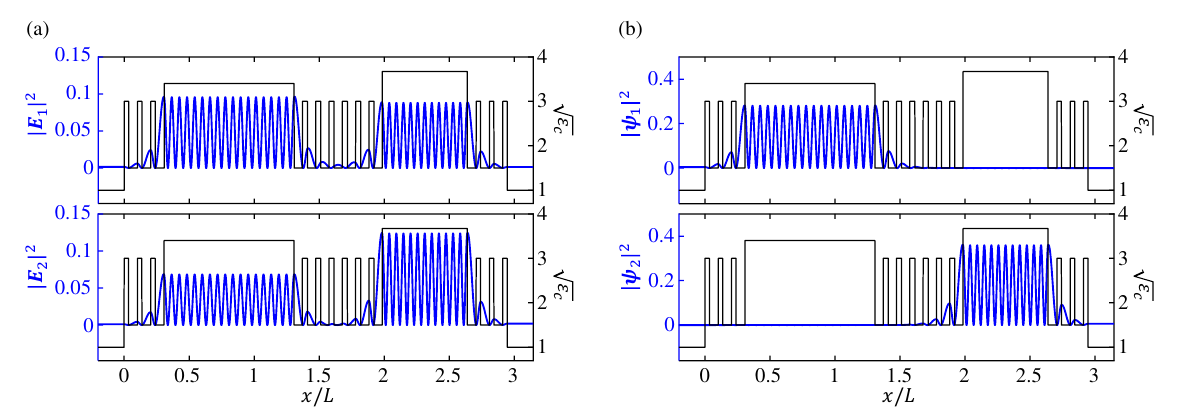}
\caption{\textbf{Basis transformation between coupled modes and local cavity modes in two coupled one-dimensional cavities.} 
\textbf{a}, Spatial intensity profiles of the system eigenmodes $|\mathbf{E}_n|^2$. 
\textbf{b}, Spatial intensity profiles of the local cavity modes $|\bm{\psi}_n|^2$, obtained as linear 
combinations of $\mathbf{E}_1$ and $\mathbf{E}_2$. The left cavity is active with length $L$, while the right 
cavity is passive.}
\label{fig:E2Psi}
\end{figure}

After normalizing $E_1$ and $E_2$ using Eq.~\eqref{orthnm_1D}, we transform the basis to local cavity modes 
$\bm{\psi}_1$ and $\bm{\psi}_2$. The transformation matrix $\mathbf{M}$ is chosen (not uniquely) such that 
$\bm{\psi}_1(x\in V_2)\approx\bm{\psi}_2(x\in V_1)\approx 0$. 
Supplementary Fig.~\ref{fig:E2Psi} illustrates the spatial profiles of the coupled-system eigenmodes and the corresponding 
local cavity modes.

\begin{figure}[h!]
\centering
\includegraphics[width=0.8\textwidth]{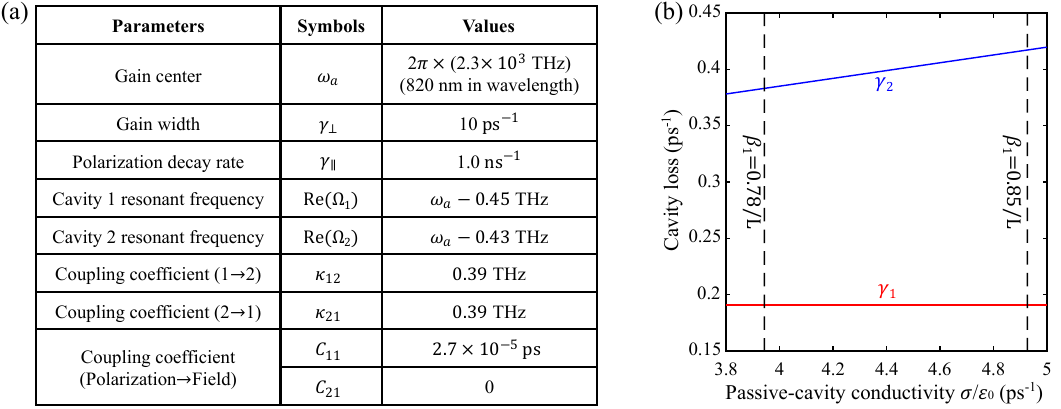}
\caption{\textbf{Parameters of the coupled-cavity system used in the main text.} 
\textbf{a}, Gain-medium and passive parameters of the two-cavity laser system. 
\textbf{b}, Dependence of the cavity loss rates $\gamma_1$ and $\gamma_2$ on the passive-cavity conductivity 
$\sigma/\varepsilon_0$. The conductivity $\sigma$ is homogeneous within the passive cavity and vanishes elsewhere. 
Black dashed lines indicate the two representative loss values used in the main text, 
$\gamma_2=0.38~\mathrm{ps}^{-1}$ and $\gamma_2=0.42~\mathrm{ps}^{-1}$.
}
\label{fig_Parameters}
\end{figure}

The elements of $\mathbf{H}_\text{L}$ are related to the material loss $\sigma(x)$ through Eqs.~\eqref{HL} and~\eqref{C}. In this example, the material loss is confined to the passive cavity and therefore contributes only to $\gamma_2$ and $C{22}$; the latter is omitted from the model because $p_2=0$. Supplementary Fig.~\ref{fig_Parameters} summarizes all B-CMT parameters used for this one-dimensional coupled-cavity system.  

\section{Stability Analysis of Fixed Points} \label{sec:Jacobian}
In this section, we derive the below-threshold Hamiltonian $\mathbf{H}_0$ in Eq.~(4) and the Bogoliubov--de Gennes (BdG) matrix in Eq.~(6) of the main text. The eigenvalue spectra of these matrices determine the linear stability of the corresponding fixed points in the laser system. A fixed point corresponds to a harmonic solution of the B-CMT equations~\eqref{BCMT:d}--\eqref{BCMT:b},
\begin{CEquation}
[d_1,~p_1,~b_1,~b_2] = [D_1,~P_1 e^{-i\omega_0 t},~B_1 e^{-i\omega_0 t},~B_2 e^{-i\omega_0 t}], \label{fixed_point}
\end{CEquation}
where $D_1$, $P_1$, and $B_{1,2}$ are time-independent amplitudes and $\omega_0\in\mathbb{R}$ is the oscillation frequency.

To analyze the stability of a fixed point, we introduce an arbitrarily small perturbation $\bm{\delta}\equiv[\delta D_1,~\delta P_1,~\delta B_1,~\delta B_2]^T$, such that
\begin{CAlign}
d_1(t) &= D_1 + \delta D_{1}(t), \\
p_1(t) &=[P_1+\delta P_{1}(t)]e^{-i\omega_0t},\\
\bm{b}(t) &= [B_1+\delta B_{1}(t), \ B_2+\delta B_{2}(t)]^\text{T} e^{-i\omega_0 t}.
\end{CAlign}
These expressions are substituted into Eqs.~\eqref{BCMT:d}--\eqref{BCMT:b} and retained to first order in $\bm{\delta}$. This linearization yields
\begin{CEquation}
i\dot{\bm{\delta}}
=
\begin{bmatrix}
-i\gamma_\parallel & \frac{2\beta_1}{\hbar}B_1^* & -\frac{2\beta_1}{\hbar}P_1^* & 0 \\
\frac{R^2}{\hbar}B_1 & \omega_a-\omega_0-i\gamma_\perp & \frac{R^2}{\hbar}D_1 & 0 \\
0 & -\frac{C_{11}}{\varepsilon_0}\omega_0^2 & \Omega_1-\omega_0 & \kappa_{12} \\
0 & -\frac{C_{21}}{\varepsilon_0}\omega_0^2 & \kappa_{21} & \Omega_2-\omega_0
\end{bmatrix}
\bm{\delta}
+
\frac{2\beta_1}{\hbar}
\begin{bmatrix}
0 & -B_1 & P_1 & 0 \\
0 & 0 & 0 & 0 \\
0 & 0 & 0 & 0
\end{bmatrix}
\bm{\delta}^*, \label{Jacobian}
\end{CEquation}
where we have applied the slowly-varying-envelope approximation, $\ddot{p}_1=-\omega_0^2p_1$.

Below the first threshold, the fixed point is trivial,
\begin{CEquation}
D_1=\nu_1,~P_1=0,~B_{1,2}=0. \label{fp_trivial}
\end{CEquation}
Under this condition, Eq.~\eqref{Jacobian} reduces to $i\dot{\bm{\delta}}=\mathbf{H}_0\bm{\delta}$, where

\begin{CEquation}
\mathbf{H}_0(\omega_0,\nu_1) =
\begin{bmatrix}
-i\gamma_\parallel & 0 & 0 & 0 \\
0 & \omega_a-\omega_0-i\gamma_\perp & \frac{R^2}{\hbar}\nu_1 & 0 \\
0 & -\frac{C_{11}}{\varepsilon_0}\omega_0^2 & \Omega_1-\omega_0 & \kappa_{12} \\
0 & -\frac{C_{21}}{\varepsilon_0}\omega_0^2 & \kappa_{21} & \Omega_2-\omega_0
\end{bmatrix},
\end{CEquation}
which is introduced as Eq.~(4) in the main text.

Above the first threshold $\nu_\text{th1}$, substituting the fixed-point ansatz in Eq.~\eqref{fixed_point} into Eqs.~\eqref{BCMT:d}--\eqref{BCMT:b} yields
\begin{CEquation}
\text{Det} [\mathbf{H}_0(\omega_0,D_1)] = 0, \label{Threshold}
\end{CEquation}
together with
\begin{CAlign}
|B_1|^2 &= \frac{\hbar^2\gamma_\perp \gamma_\parallel}{4\beta_1R^2|\Gamma_\perp(\omega_0)|^2} \left(\frac{\nu_1}{D_1}-1\right), \label{fp_B} \\
P_1 &= \frac{R^2}{\hbar\gamma_\perp}\Gamma_\perp(\omega_0){D}_1B_1, \label{fp_P}
\end{CAlign}
where $\Gamma_\perp(\omega)=\gamma_\perp/(\omega-\omega_a+i\gamma_\perp)$. Equation~\eqref{Threshold} is a complex scalar equation for the two real variables $\omega_0$ and $D_1$, independent of $\nu_1$. In general, multiple solutions may exist; the physically relevant first lasing mode corresponds to the solution with the smallest $D_1$. The steady-state field intensity and polarization are then related to the pump $\nu_1$ through Eqs.~\eqref{fp_B} and~\eqref{fp_P}. Due to the $U(1)$ symmetry of the B-CMT equations, the phase of $B_1$ is arbitrary, allowing us to choose the gauge $B_1=|B_1|$. At threshold, the nontrivial fixed point must continuously connect to the trivial one in Eq.~\eqref{fp_trivial}, which yields the threshold pump value $\nu_\text{th1}=D_1$.

With these relations, substituting Eq.~\eqref{fp_P} into Eq.~\eqref{Jacobian} gives
\begin{CEquation}
i\dot{\bm{\delta}} = (\mathbf{H}_0+\frac{B_1}{\hbar}\bm{\Delta})\bm{\delta}+\frac{B_1}{\hbar}\mathbf{K}\bm{\delta}^*, \label{hBDG}
\end{CEquation}
where
\begin{CAlign}
\bm{\Delta} =
\begin{bmatrix}
0 & {2\beta_1} & -\frac{2\beta_1R^2}{\hbar\gamma_\perp}\Gamma_\perp^*(\omega_0)\nu_\text{th1} & 0 \\
R^2 & 0 & 0 & 0\\
0 & 0 & 0 & 0\\
0 & 0 & 0 & 0
\end{bmatrix}, ~
\bm{K}=
\begin{bmatrix}
0 & -{2\beta_1} & \frac{2\beta_1R^2}{\hbar\gamma_\perp}\Gamma_\perp(\omega_0)\nu_\text{th1} & 0 \\
0 & 0 & 0 & 0\\
0 & 0 & 0 & 0\\
0 & 0 & 0 & 0
\end{bmatrix}. \label{LIH}
\end{CAlign}
Finally, combining Eq.~\eqref{hBDG} with its complex conjugate, we rewrite the perturbation theory in the form of a standard linear Schr\"{o}dinger equation,
\begin{CEquation}
i\frac{d}{dt}
\begin{bmatrix}
\bm{\delta} \\ \bm{\delta}^*
\end{bmatrix}
= \bigg(
\begin{bmatrix}
\mathbf{H}_0(\nu_{\mathrm{th}1}) & \mathbf{0} \\
\mathbf{0} & -\mathbf{H}_0^*(\nu_{\mathrm{th}1})
\end{bmatrix}
+\frac{B_1}{\hbar}
\begin{bmatrix}
\mathbf{\Delta} & \mathbf{K} \\
-\mathbf{K}^* & -\mathbf{\Delta}^*
\end{bmatrix}
\bigg)
\begin{bmatrix}
\bm{\delta} \\ \bm{\delta}^*
\end{bmatrix}
\equiv
\mathbf{H}_\text{BdG}
\begin{bmatrix}
\bm{\delta} \\ \bm{\delta}^*
\end{bmatrix}. \label{HBdG}
\end{CEquation}
Equation~\eqref{HBdG} corresponds to Eq.~(6) in the main text, where we have fixed the gauge such that $B_1=|B_1|=|d_1|$.

Let $\tilde{\omega}$ be an eigenvalue of $\mathbf{H_\text{BdG}}$ and $\bm{q}(\tilde{\omega})=[q_1,q_2,\dots,q_8]^T$ the corresponding eigenvector. Since the population-inversion fluctuation $\delta D_1$ must be real, physical eigenvectors are constrained by
\begin{CEquation}
({q}_1 + {q}_5^*) \in \mathbb{R}. \label{D-condition}
\end{CEquation}
The eigenvalue $\tilde{\omega}=0$ corresponds to the $\mathfrak{u}(1)$ transformation of the lasing mode, with eigenvector $\bm{q}\propto i[0,P_1,B_1,B_2,0,-P_1^*,-B_1^*,-B_2^*]$, which automatically satisfies Eq.~\eqref{D-condition}.

For $\tilde{\omega}\neq 0$ and $\nu_1>\nu_\text{th1}$, we apply an isospectral transformation to $\mathbf{H}_\text{BdG}$, yielding
\begin{CEquation}
\bigg(\frac{B_1R}{\hbar}\bigg)^2
\begin{bmatrix}
X(\tilde{\omega}) & Y(\tilde{\omega}) \\ X(\tilde{\omega}) & Y(\tilde{\omega})
\end{bmatrix}
\begin{bmatrix}
q_1 \\ q_5
\end{bmatrix}
=(\tilde{\omega}+i\gamma_\parallel)
\begin{bmatrix}
q_1 \\ q_5
\end{bmatrix}, \label{isotrans}
\end{CEquation}
where
\begin{CAlign}
X(\tilde{\omega})
& =
\begin{bmatrix}
0 & -{2\beta_1} & \frac{2\beta_1R^2}{\hbar\gamma_\perp}\Gamma_\perp^*(\omega_0)\nu_\text{th1} & 0
\end{bmatrix}
[\mathbf{H}_0(\nu_\text{th1})-\tilde{\omega} \mathbf{I}]^{-1}
\begin{bmatrix}
0 & 1 & 0 &0
\end{bmatrix}^T, \\
Y(\tilde{\omega})
& =
\begin{bmatrix}
0 & -{2\beta_1} & \frac{2\beta_1R^2}{\hbar\gamma_\perp}\Gamma_\perp(\omega_0)\nu_\text{th1} & 0
\end{bmatrix}
[\mathbf{H}_0^*(\nu_\text{th1})-\tilde{\omega} \mathbf{I}]^{-1}
\begin{bmatrix}
0 & 1 & 0 &0
\end{bmatrix}^T.
\end{CAlign}
The eigenvalue $\tilde{\omega}=-i\gamma_\parallel$ trivially satisfies Eq.~\eqref{isotrans}. However, the associated eigenvector obeys $q_1/q_5=-Y/X$, which does not, in general, satisfy the reality condition in Eq.~\eqref{D-condition} and therefore corresponds to a non-physical inversion fluctuation.

All remaining eigenvalues must instead satisfy
\begin{CEquation}
\tilde{\omega}+i\gamma_\parallel=\bigg(\frac{B_1R}{\hbar}\bigg)^2[X(\tilde{\omega})+Y(\tilde{\omega})],
\end{CEquation}
which enforces $q_1=q_5$ and guarantees that the resulting inversion fluctuation is real. These eigenvalues therefore correspond to physical BdG modes discussed in the main text.

\section{Quartic-Equation Approximation for the $\mathbf{H}_{\mathrm{BdG}}$ Eigenvalue Spectrum}
The system discussed in the main text operates close to, but not exactly at, an exceptional point when $\nu_1=\nu_\text{th1}$. As a result, $\mathbf{H}_0(\nu_\text{th1})$ remains diagonalizable and can be written as
\begin{CEquation}
\mathbf{W}^{-1}\mathbf{H}_0(\nu_\text{th1})\mathbf{W}=\Lambda_0\equiv
\begin{bmatrix}
0 & 0 & 0 & 0 \\
0 & \omega_\text{II} & 0 & 0 \\
0 & 0 & \omega_\text{III} & 0 \\
0 & 0 & 0 & -i\gamma_\parallel
\end{bmatrix}.
\end{CEquation}
Here $\mathbf{W}$ is the $4\times4$ matrix formed by the right eigenvectors of $\mathbf{H}_0(\nu_\text{th1})$, while $\mathbf{W}^{-1}$ contains the corresponding left eigenvectors. We label the columns of $\mathbf{W}$ and the rows of $\mathbf{W}^{-1}$ using bra–ket notation as
\begin{CEquation}
    \mathbf{W}=
    \begin{bmatrix}
     \ket{\text{R}_\text{I}} & \ket{\text{R}_\text{II}} & \ket{\text{R}_\text{III}} & \ket{\text{I}} 
    \end{bmatrix},~
    \mathbf{W}^{-1}=
    \begin{bmatrix}
     \bra{\text{L}_\text{I}} \\ \bra{\text{L}_\text{II}} \\ \bra{\text{L}_\text{III}} \\ \bra{\text{I}} 
    \end{bmatrix}. \label{W}
\end{CEquation}
The kets $\ket{\text{R}_\text{I}}$ and $\ket{\text{R}_\text{II}}$ correspond to the two optical modes, $\text{O}_\text{I}$ and $\text{O}_\text{II}$ defined in the main text. The eigenvalue $\omega_\text{II}$ denotes the eigenfrequency bias of $\text{O}_\text{II}$ evaluated at the threshold, i.e., $\omega_\text{II}=\omega_\text{II}(\nu_\text{th1})$. The third eigenvalue $\omega_\text{III}$ lies near the singularity of $\Gamma_\perp(\omega+\omega_0)$ and does not play a role in the relevant dynamics. The fourth eigenvalue $-i\gamma_\parallel$ represents the decay of population-inversion fluctuations. The right eigenvector $\ket{\text{R}_\text{I}}$, together with the inversion eigenvector $\ket{\text{I}}$ and its dual $\bra{\text{I}}$, can be written analytically in terms of the fixed-point solution (unnormalized) as
\begin{CEquation}
    \ket{\text{R}_\text{I}} = 
    \begin{bmatrix}
        0 & P_1 & B_{1} & B_2
    \end{bmatrix}^T,~
    \ket{\text{I}} = (\bra{\text{I}})^T = 
        \begin{bmatrix}
        1 & 0 & 0 & 0
    \end{bmatrix}^T. \label{OI}
\end{CEquation}

We now compute the eigenvalues of $\mathbf{H}_\text{BdG}$ in the eigenbasis defined by $\mathbf{W}$. Performing the associated similarity transformation, the characteristic equation becomes,
\begin{CEquation}
    \text{Det}\bigg(
    \begin{bmatrix}
      \mathbf{W} & \mathbf{0} \\ \mathbf{0} & \mathbf{W}^{*}  
    \end{bmatrix}^{-1}
    \mathbf{H}_\text{BdG}
    \begin{bmatrix}
      \mathbf{W} & \mathbf{0} \\ \mathbf{0} & \mathbf{W}^{*}  
    \end{bmatrix}
    -\tilde{\omega}\mathbf{I}
    \bigg)
    =
    \text{Det}\bigg(
    \begin{bmatrix}
        \Lambda_0 & \bm{0} \\
        \bm{0} & -\Lambda_0^*
    \end{bmatrix}
    -\tilde{\omega} \mathbf{I}+
    \frac{B_1}{\hbar}\Lambda^\text{LIH}
    \bigg)=0, \label{BdG_eigen}
\end{CEquation}
where the LIH coupling matrix in the transformed basis is defined as
\begin{CEquation}
    \Lambda^\text{LIH}=
    \begin{bmatrix}
      \mathbf{W} & \mathbf{0} \\ \mathbf{0} & \mathbf{W}^{*}  
    \end{bmatrix}^{-1}
    \begin{bmatrix}
\mathbf{\Delta} & \mathbf{K} \\
-\mathbf{K}^* & -\mathbf{\Delta}^*
\end{bmatrix}
\begin{bmatrix}
      \mathbf{W} & \mathbf{0} \\ \mathbf{0} & \mathbf{W}^{*}  
\end{bmatrix} =
\begin{bmatrix}
    \mathbf{W}^{-1}\bm{\Delta}\mathbf{W} & \mathbf{W}^{-1}\mathbf{K}\mathbf{W}^* \\
    -(\mathbf{W}^{-1}\mathbf{K}\mathbf{W}^*)^* & -(\mathbf{W}^{-1}\bm{\Delta}\mathbf{W})^*
\end{bmatrix}.
\label{LIH_Trans}
\end{CEquation}
This transformation preserves the Bogoliubov–de Gennes symmetry of the spectrum. Substituting Eqs.~\eqref{LIH}, \eqref{W}, and \eqref{OI} into Eq.~\eqref{LIH_Trans}, we identify the nonzero elements of $\Lambda^\text{LIH}$ as
\begin{CEquation}
\begin{aligned}
&\Lambda^\text{LIH}_{41}=\Lambda^\text{LIH}_{81}=-(\Lambda^\text{LIH}_{45})^*=-(\Lambda^\text{LIH}_{85})^*=\bra{\text{I}}\bm{\Delta}\ket{\text{R}_\text{I}}\equiv \beta_1S_1, \\
&\Lambda^\text{LIH}_{42}=\Lambda^\text{LIH}_{82}=-(\Lambda^\text{LIH}_{46})^*=-(\Lambda^\text{LIH}_{86})^*=\bra{\text{I}}\bm{\Delta}\ket{\text{R}_\text{II}}\equiv \beta_1 S_2, \\
&\Lambda^\text{LIH}_{43}=\Lambda^\text{LIH}_{83}=-(\Lambda^\text{LIH}_{47})^*=-(\Lambda^\text{LIH}_{87})^*=\bra{\text{I}}\bm{\Delta}\ket{\text{R}_\text{III}}\equiv \beta_1 S_3,   \\
&\Lambda^\text{LIH}_{14}=-(\Lambda^\text{LIH}_{58})^*=\bra{\text{L}_\text{I}}\bm{\Delta}\ket{\text{I}}\equiv R^2 S_4, \\
&\Lambda^\text{LIH}_{24}=-(\Lambda^\text{LIH}_{68})^*=\bra{\text{L}_\text{II}}\bm{\Delta}\ket{\text{I}}\equiv R^2 S_5, \\
&\Lambda^\text{LIH}_{34}=-(\Lambda^\text{LIH}_{78})^*=\bra{\text{L}_\text{III}}\bm{\Delta}\ket{\text{I}}\equiv R^2 S_6. 
\end{aligned} \label{S}
\end{CEquation}
With these definitions, the characteristic equation \eqref{BdG_eigen} equals to 
\begin{CEquation}
\begin{aligned}
    &\tilde{\omega}^2(\tilde{\omega}+i\gamma_\parallel)^2(\tilde{\omega}-\omega_\text{II})(\tilde{\omega}+\omega_\text{II}^*)(\tilde{\omega}-\omega_\text{III})(\tilde{\omega}+\omega_\text{III}^*) \\
    &-\frac{R^2\beta_1B_1^2}{\hbar^2}\tilde{\omega}^2(\tilde{\omega}+i\gamma_\parallel)(\tilde{\omega}-\omega_\text{II})(\tilde{\omega}+\omega_\text{II}^*)[S_3^*S_6^*(\tilde{\omega}-\omega_\text{III})+S_3S_6(\tilde{\omega}+\omega_\text{III}^*)]\\
    &-\frac{R^2\beta_1B_1^2}{\hbar^2}\tilde{\omega}^2(\tilde{\omega}+i\gamma_\parallel)(\tilde{\omega}-\omega_\text{III})(\tilde{\omega}+\omega_\text{III}^*)[S_2^*S_5^*(\tilde{\omega}-\omega_\text{II})+S_2S_5(\tilde{\omega}+\omega_\text{II}^*)] \\
    &-\frac{R^2\beta_1B_1^2}{\hbar^2}\tilde{\omega}(\tilde{\omega}+i\gamma_\parallel)(\tilde{\omega}-\omega_\text{II})(\tilde{\omega}+\omega_\text{II}^*)(\tilde{\omega}-\omega_\text{III})(\tilde{\omega}+\omega_\text{III}^*)[(S_1-S_1^*)(S_4-S_4^*)] \\
    =&0. \label{BdG_Poly}
\end{aligned}
\end{CEquation}
The first line of Eq.~\eqref{BdG_Poly} reproduces the eigenvalue structure of $\mathbf{H}\text{BdG}$ at the first threshold, where both $\tilde{\omega}=0$ and $\tilde{\omega}=-i\gamma\parallel$ exhibit twofold degeneracy. Above threshold, the lasing intensity $B_1^2$ becomes nonzero. While $\tilde{\omega}=0$ and $\tilde{\omega}=-i\gamma_\parallel$ remain exact solutions, their degeneracies are lifted, consistent with the numerical results shown in Fig.~2\textbf{a} of the main text.

Since $\text{Im}(\omega_\text{III})\approx-\gamma_\perp$ is much larger in magnitude than $\text{Im}(\omega_\text{II})$ and $-\gamma_\parallel$, two of the solutions of Eq.~\eqref{BdG_Poly} remain near $\omega_\text{III}$ and $-\omega_\text{III}^*$ and are well separated from the remaining eigenvalues. Excluding these two solutions implies $(\tilde{\omega}-\omega_\text{III}),(\tilde{\omega}+\omega_\text{III}^*)\gg\tilde{\omega},(\tilde{\omega}+i\gamma_\parallel),(\tilde{\omega}-\omega_\text{II}),(\tilde{\omega}+\omega_\text{II}^*)$. 
In this regime, the second-row terms in Eq.~\eqref{BdG_Poly}, which contain either $(\tilde{\omega}-\omega_\text{III})$ or $(\tilde{\omega}+\omega_\text{III}^*)$, are negligible compared to the remaining contributions proportional to $(\tilde{\omega}-\omega_\text{III})(\tilde{\omega}+\omega_\text{III}^*)$. Consequently, for the eigenvalues relevant to lasing dynamics, Eq.~\eqref{BdG_Poly} can be approximated by
\begin{CEquation}
    Q(\tilde{\omega},B_1^2) \approx 0,
\end{CEquation}
where $Q(\omega,B_1^2)$ is a quartic function of $\omega$,
\begin{CEquation}
    Q(\omega,B_1^2) = 
    \omega(\omega+i\gamma_\parallel)(\omega-\omega_\text{II})(\omega+\omega_\text{II}^*)-
    S_v(B_1^2)(s_2\omega^2-is_1\omega-s_0), \label{Q}
\end{CEquation}
whose coefficients are given by
\begin{CEquation}
\begin{aligned}
     s_v & = 2R^2\beta_1B_1^2{\hbar^{-2}}, \\
     s_0 &= -2|\omega_\text{II}|^2\text{Im}(S_1)\text{Im}(S_4), \\
     s_1 &= \text{Im}(S_2^*S_5^*\omega_\text{II})-4\text{Im}(\omega_\text{II})\text{Im}(S_1)\text{Im}(S_4), \\
     s_2&=\text{Re}(S_2S_5)-2\text{Im}(S_1)\text{Im}(S_4).
\end{aligned}
\end{CEquation}
Finally, adopting the gauge choice $B_1=|B_1|$ from Sec.~\ref{sec:Jacobian} allows us to substitute Eq.~\eqref{fp_B} into Eq.~\eqref{Q}, yielding Eq.~(9) in the main text.

We denote the four roots of $Q$ as $\tilde{\omega}_\text{I}$, $\tilde{\omega}_\text{II}$, and their particle–hole–symmetric counterparts. An exceptional point occurs when the two LIH branches coalesce, $\tilde{\omega}_\text{I}=\tilde{\omega}_\text{II}\equiv \omega_\text{EP}$, which leads to Eq.~(10) in the main text,
\begin{CEquation}
    Q(\omega)=(\omega-\omega_{\mathrm{EP}})^2(\omega+\omega_{\mathrm{EP}}^{*})^2. \label{LIH-EP}
\end{CEquation}
Equating the polynomial coefficients in Eq.~\eqref{Q} and Eq.~\eqref{LIH-EP} yields the following set of constraints,
\begin{CAlign}
    \text{Im}(\omega_\text{EP}) &= -\frac{h_3}{4}, \label{LIH_1}\\
    2|\omega_\text{EP}|^2-s_v s_2 &= h_2-\frac{h_3^2}{4}, \label{LIH_2}\\
    |\omega_\text{EP}|^4 + s_v s_0 &= 0, \label{LIH_3}\\
    4|\omega_\text{EP}|^2\text{Im}(\omega_\text{EP}) - s_vs_1 &= -h_1, \label{LIH_4}
\end{CAlign}
where 
\begin{CAlign}
  h_1 &= \gamma_\parallel |\omega_\text{II}|^2, \\
  h_2 &= |\omega_\text{II}|^2 - 2\gamma_\parallel \text{Im}(\omega_\text{II}), \\
  h_3 &= \gamma_\parallel - 2 \text{Im} (\omega_\text{II}).
\end{CAlign}
By combining Eqs.~\eqref{LIH_1}–\eqref{LIH_4}, we eliminate both $s_v$ and $\omega_\text{EP}$ and obtain the condition
\begin{CEquation}
    1 = \frac{s_0(0.25h_3^3+2h_1-h_2h_3)(2s_1+s_2h_3)}{(h_1s_2+h_2s_1-0.25h_3^2s_1)^2} \equiv \zeta. \label{zeta}
\end{CEquation}
This dimensionless parameter $\zeta$ characterizes the occurrence of the LIH-induced exceptional point.

In slow-gain semiconductor lasers where the inversion relaxation rate is small compared to the detuning of the nonlasing optical mode, such as the example considered in Sec.~\ref{1D-laser}, the condition $\text{Im}(\omega_\text{II}) \ll -\gamma_\parallel$ holds. In this limit, we may approximate $h_1 \approx 0$, $h_2 \approx |\omega_\text{II}|^2$, and $h_3 \approx -2,\text{Im}(\omega_\text{II})$. Under these conditions, Eq.~\eqref{zeta} simplifies directly to Eq.~(11) in the main text.

\bibliographystyle{naturemag}
\bibliography{sn-bibliography}